\documentclass[sn-mathphys-ay]{sn-jnl}

\usepackage{graphicx}%
\usepackage{multirow}%
\usepackage{amsmath,amssymb,amsfonts}%
\usepackage{amsthm}%
\usepackage{mathrsfs}%
\usepackage[title]{appendix}%
\usepackage{xcolor}%
\usepackage{textcomp}%
\usepackage{manyfoot}%
\usepackage{booktabs}%
\usepackage{algorithm}%
\usepackage{algorithmicx}%
\usepackage{algpseudocode}%
\usepackage{listings}%

\usepackage{anyfontsize}

\newcommand{\degree}{^\circ{}}

\begin{document}

\title[The Asteroid Framing Cameras on ESA’s Hera mission]{The Asteroid Framing Cameras on ESA’s Hera mission}


\author*[1]{  \fnm{Jean-Baptiste} \sur{Vincent}}\email{jean-baptiste.vincent@dlr.de}
\author[2]{   \fnm{Gábor}         \sur{Kovács}}
\author[2]{   \fnm{Balázs V.}     \sur{Nagy}}
\author[1]{   \fnm{Frank}         \sur{Preusker}}
\author[3]{   \fnm{Naomi}         \sur{Murdoch}}
\author[4]{   \fnm{Maurizio}      \sur{Pajola}}
\author[5]{   \fnm{Michael}       \sur{Kueppers}}
\author[6,7]{ \fnm{Patrick}       \sur{Michel}}
\author[7]{   \fnm{Seiji}         \sur{Sugita}}
\author[8]{   \fnm{Hannah}        \sur{Goldberg}}

\affil*[1]{\orgname{DLR Institute of Planetary Research}, \city{Berlin}, \country{Germany}}
\affil[2]{ \orgname{Budapest University of Technology and Economics, Department of Mechatronics, Optics and Engineering Informatics}, \city{Budapest}, \country{Hungary}}
\affil[3]{ \orgname{Institut Supérieur de l'Aéronautique et de l'Espace (ISAE-SUPAERO), Université de Toulouse}, \city{Toulouse}, \country{France}}
\affil[4]{ \orgname{INAF, Astronomical Observatory of Padova}, \city{Padova}, \country{Italy}}
\affil[5]{ \orgname{European Space Agency, European Space Astronomy Centre}, \city{Villanueva de la Cañada, Madrid}, \country{Spain}}
\affil[6]{ \orgname{Université Côte d'Azur, Observatoire de la Côte d'Azur, CNRS, Laboratoire Lagrange}, \city{Nice}, \country{France}}
\affil[7]{ \orgname{The University of Tokyo, Department of Systems Innovation, School of Engineering}, \city{Tokyo},  \country{Japan}}
\affil[8]{ \orgname{University of California, Berkeley Space Science Laboratory}, \city{Berkeley},  \country{USA}}

\abstract{As the first asteroid deflection test, NASA’s successfully hit asteroid Dimorphos (secondary of the binary asteroid 65803 Didymos) with the DART kinetic impactor on September 26, 2022. To fully characterise the physical properties of the objects, and measure precisely the effects of this impact in the context of planetary defence, ESA launched the Hera mission on 7 October 2024, with scheduled arrival at Didymos in fall 2026.
Among the core payload of the mission, the Asteroid Framing Cameras are two identical imaging systems that will support navigation and scientific activities, by acquiring images from various distances and observing geometries during the course of the mission.
Built by Jena-Optronik (Germany), the cameras match the requirements designed by the science team and will provide data that supports a wide range of investigations: hazard detection, system dynamics, shape reconstruction, surface morphology and mapping, and surface photometry.
Each instrument is a panchromatic camera equipped with a 5.5 x 5.5 degree field of view, and an angular resolution of 93.7 micro-radians per pixel. The cameras shall provide the necessary data to address the mission requirements through a global mapping of the two components of the binary system at spatial scales of 2–3 m/pixel in the Early Characterisation Phase, 1–2 m/pixel in the Detailed Characterisation Phase, and 0.5-2 m/pixel in the Close Operation Phase. Dedicated flybys will bring the resolution down to $<$ 10 cm/pixel on specific areas of interest on Dimorphos, such as the DART impact site and the JUVENTAS cubesat landing site.
Here, we present the technical specifications of the camera, as well as the status of the calibration. We then summarise the planned operations in cruise and at the asteroids. Finally, we provide examples of the scientific investigations and products that will make use of the data returned by the cameras.
}




\maketitle

\section{Introduction}\label{sec:intro}
Hera is the European Space Agency (ESA) component of the large international project AIDA: Asteroid Impact and Deflection Assessment. The mission is designed to study the effects of a kinetic impact on an asteroid in the context of international Planetary Defence activities \citep{michel2022}.  More specifically, Hera will assess the outcome of NASA's Double Asteroid Redirection Test (DART, \cite{cheng2018}) which intentionally collided with the small moon called Dimorphos of the binary asteroid (65803) Didymos on 26 September 2022. DART achieved its objective of hitting the target asteroid at more than 6 km/s \citep{daly2023}, and created a measurable decrease of its orbital period around the larger parent asteroid Didymos \citep{thomas2023}. However, a true assessment of the efficiency of the deflection, necessary for designing future responses to actual hazardous asteroids, requires a dedicated mission for in-depth characterisation. That is exactly the purpose of Hera. After launch on 7 October 2024, Hera will reach the Didymos system in late 2026 with the following goals:
\begin{itemize}
    \item determine dynamical properties of the Didymos system: semi-major axis, eccentricity,
spin rates and orientation
    \item measure global physical properties of each asteroid; including internal ones: mass, volume, density, size
distribution of surface material
    \item characterise the shape of the DART crater, and/or extent of large scale reshaping
    \item assess the presence of excavated material on the surface or in the system and determine its size distribution
\end{itemize}

These primary objectives are necessary to fully establish the momentum transfer imparted by DART (a preliminary assessment can be found in \cite{cheng2023}), and thus close the AIDA experiment. A more detailed description of the mission, including all requirements and a science traceability matrix can be found in \cite{michel2022}.

To achieve these goals, Hera carries a selection of payload carefully chosen to meet the primary requirements of the mission. In this article, we focus on the Asteroid Framing Cameras, a set of two identical panchromatic imaging instruments to be used as the core payload for navigation and asteroid characterisation. In the following sections, we will introduce the instruments and their technical specifications, the high-level operational plan for the mission, and the calibration measurements that have been acquired before launch. We will conclude with an overview of the expected scientific return to be enabled by the AFCs, which extends beyond the Planetary Defence requirements of the mission, due to the unique nature of the Didymos system and the many open questions brought forward by DART (e.g. see morphological considerations in \cite{barnouin2024}

\section{Requirements and Specifications}\label{sec:tech_desc}
The Asteroid Framing Cameras (AFC) are a core payload of Hera, necessary for both navigation and  completion of the mission objectives. As a risk mitigation measure, the mission flys with two identical cameras, AFC1 and AFC2, which have the exact same specifications and performance. Both cameras are totally interchangeable for all intents and purposes. During the mission, all calibration operations are implemented and executed identically for both instruments. However, Science/Navigation data is acquired exclusively with AFC1, and AFC2 is reserved for contingency mitigation.

The AFC are panchromatic cameras with refractive optics, operating in the visible wavelength range (400-900 nm). Provided by Jena-Optronik, they are based on the ASTROhead design\footnote{\url{https://www.jena-optronik.de/products/cameras-and-camera-systems/astrohead-cam.html}, retrieved on 2025-11-03}, and use a FaintStar2 CMOS detector \footnote{\url{https://caeleste.be/news/eyes-on-hera-with-faintstar2-detector-chip/}, retrieved on 2025-11-03}. A detailed list of the cameras specifications is provide in Table \ref{tab:specs}. These cameras have the highest Technology Readiness Level (TRL9) , having successfully flown on multiple missions. The first ASTROhead Cam flight set was used on Northrop Grumman’s Mission Extension Vehicle MEV-1 in 2019, which made a rendez-vous with commercial satellite Intelsat 901 on 25 February 2020. A similar mission MEV-2, docked with Intelsat 10-02 on a geostationary orbit on 12 April 2021, once again supported by a set of ASTROhead cameras.

Besides the high TRL, the AFC were selected because they fulfil all mission requirements for an imaging instrument. The most relevant constraints are the following (from ESA's Hera Mission Requirements document)\footnote{\url{https://www.heramission.space/missionrequirementsdocuments}, retrieved on 2025-11-03}:

\begin{enumerate}
\item Hera shall identify the DART impact area on Dimorphos and acquire data to reconstruct the tri-dimensional shape of the eventual crater to an accuracy of 50 cm. Because of the inherent loss of resolution from 3D reconstruction techniques, this requires an image pixel scale of 10 cm (one needs to correlate multiple images of the same region with varying viewing geometry and the matching of features between image pairs require multiple pixels). For safety, Hera remains outside of the orbit of Dimorphos and does not approach closer than 1 km from either asteroid during the nominal mission. At this distance, the 10 cm/px requirement constrains the angular scale (a.k.a. IFoV: Instantaneous Field-of-View) to be 100 $\mu$rad/px or smaller.

\item Hera shall determine the mass of Dimorphos within 10\%. This can be achieved by determining the motion of Didymos around the common centre of mass of Didymos and Dimorphos to an accuracy of 1 m, by tracking landmarks on the surface of the asteroids. To guarantee a successful tracking, it is preferable that the camera can observe the full body of Didymos during several orbits of Dimorphos,  with a pixel scale of 1 meter. The previous requirement already constrains the angular resolution; from which we know that a pixel scale of 1 m can be obtained from a distance of 10 km. At this distance, the angular size of Didymos (max diameter 850 m) is 4.85$\degree$, which sets the minimum size of the camera field-of-view.
\end{enumerate}

Therefore, the minimum set of constraints for an imaging instrument on Hera are IFOV $<= 100 \mu rad/px$ and FOV $>= 4.85\degree$. Table \ref{tab:specs} shows that the AFC surpasses both requirements.

\begin{table}[h]
\caption{Specifications (from manufacturer)}\label{tab:specs}%
\begin{tabular}{l|l}
\toprule
Parameter           & Value                    \\
\midrule
Field of View       & $5.5 \times 5.5\degree$  \\
Image size          & $1020 \times 1020$ pixel \\
Focal length        & 105.57 mm                \\
F-number            & 4.2                      \\
Angular scale       & 94.1 $\mu$rad/px         \\
Spectral range      & 400-900 nm               \\
A/D resolution      & 12 bits (4096 values)    \\
ADC conversion      & 32 $e^-/DN$              \\
Integration time    & 224 $\mu$s to 5 s        \\
Readout frequency   & Up to 4 Hz               \\
Data volume / image & 1.5 MByte                \\
Mass                & 1.1 kg                   \\
Power               & 0.9 W                    \\
\botrule
\end{tabular}
\end{table}

\section{Calibration}\label{sec:calib}
The AFC is a space certified instrument designed for navigation. For scientific use, one needs to ensure a thorough calibration on-ground and in-flight, in order to convert reliably the pixel values (Digital Numbers, a.k.a. DN) into physical quantities (e.g. asteroid reflectance). Here we describe the methods and results of the Hera AFC ground calibration. The optical tests and the image acquisitions have been performed in December 2022 and January 2023, at Jena Optronik (JOP) in Jena, Germany. All tasks were based on recommendations from the Hera Investigation Team, and follow procedures developed for imaging instruments on previous missions (i.e. \cite{kovacs2024}). The analysis and the evaluation of the results have been performed at the Budapest University of Technology and Economics, Department of Mechatronics, Optics and Engineering Informatics. The procedure and results have been evaluated and validated by the Hera Investigation Team, within the frame work of the "Hera Data Analysis, Exploitation, Interpretation Working Group" (a.k.a. WG4).

The calibration tests were carried out with the same settings as currently planned for the scientific imaging phases of the mission:
\begin{itemize}
\item Full field of view (1020x1020 pixels)
\item 12 bits ADC setting, providing 4096 DN levels for each image
\item No image compression was used
\item Commanded exposure time was set between 0.224 ms and 5000.0 ms, covering the full range of capabilities of the instrument
\item The acquired images were stored as standard FITS image files. The FITS header contains the camera settings, exposure time, temperatures, and other information about the instruments and the actual operation.
\end{itemize}

In the following subsections, we detail the calibration methods and results related to on-ground measurement of dark, bias, flat fields, linearity, and optical distortion.

\subsection{Dark \& Bias}
The temperature dependency dark test has been carried out by the JOP personnel in December 2022, in a thermal vacuum chamber at the JOP optical laboratory. The tests resulted in FITS data cubes at various temperatures and exposure times. The nominal (interface plate) temperatures were set to -25°C, -15°C, 0°C, 10°C, 20°C, 35°C and 45°C. The actual detector temperatures are sometimes slightly higher due to the interface plate in the setup, but the actual sensor temperature was recorded and stored in the FITS header. After stabilising the system at each temperature, image cubes were acquired with 0.224 ms, 100 ms, 500 ms, 1000 ms, and 5000 ms exposure times. Each cube consists of 30 images acquired with the same exposure and detector temperature.
During the tests, the camera aperture was covered by a black cup, and the laboratory lights were switched off. The ambient temperature was also recorded.

The camera does not have a shutter, therefore a true 0 ms exposure is not possible. For this reason, the bias frame is approximated by the shortest possible dark exposure time, which was 0.224 ms. The maximum dark exposure time was 5000 ms.

General findings:
\begin{itemize}
\item The bias images showed the expected uniform noise pattern in the $\pm 2$ DNs range.
\item The individual bias images did not exhibit any structure or pattern noise, only random noise was visible.
\item The average bias value was stable throughout the full test campaign within $\pm 1$ DN.
\item The bias images exhibited a slight temperature dependency in the fraction of a DN. (it might be the effect of the finite exposure time)
\item The short exposure dark images showed the expected uniform noise pattern.
\item The exposure time increase resulted in the expected linear DN increase.
\item The 1000 ms and 5000 ms dark images exhibited a slight increase in the background values towards the detector edges, which might be an effect of a chip temperature gradient. The average increase was 11 DNs at $35\degree C$ interface temperature, and 5000 ms exposure.
\end{itemize}

For each camera, a master bias frame for the calibration pipeline was created from the $20\degree C$ interface temperature image cube. The FITS cube contains 30 consecutive images with 0.224 ms exposure time. Each pixel of the reference bias frame is calculated as the median of the same coordinate 30 pixels. This median bias image is stored as a single frame FITS image for each camera as: “AFC1\_BIAS\_V1.0.FTS” and “AFC2\_BIAS\_V1.0.FTS” and used for the calculations in the data processing of all calibration measurements.

Similarly, master dark frames for the calibration pipeline were created from the $20\degree C$ interface temperature image cube. The FITS cube contains 30 consecutive images with 999.904 ms exposure time. Each pixel of the reference dark frame is calculated as the median of the same coordinate 30 pixels. This median dark image is stored as a single frame FITS image for each camera as: “AFC1\_DARK\_V1.0.FTS” and “AFC2\_DARK\_V1.0.FTS” and used for the calculations in the data processing of all calibration measurements. The actual detector temperature of the master dark frame is recorded in the image header and should be used during the scaling of the dark frame.

The dark current $D$ (in DN/s) of a typical pixel follows the formula:
\[
D(T) = A e^{\frac{-B}{k_BT}}
\]
where $A = 8.91\times10^{12}$, $B = 1.02\times10^{-19}$ are instrument-specific constants, $T$ is the detector temperature in Kelvin, and $k_B$ is Boltzmann's constant ($1.38\times10^{-23} m^2 kg s^{-2} K^{-1}$).

In practical terms, the change of temperature needed to double the dark current is given by:
\[
\frac{1}{T_2} = \frac{1}{T_1} - \frac{k_B \ln{2}}{B}
\]

with temperatures expressed in Kelvin.

For the reference temperature $T_1 = 20.0^{\circ}C$ (293.15 K), we get $T_2 = 28.28^{\circ}C$ (301.28 K), i.e. an increase of $8.13^{\circ}C$.

The scaling factor for the reference dark frame is given by:
\[
C_{dark} = \frac{t_{exp}}{t_{expDark}} e^{\left(\frac{-B}{k_B}\left(\frac{1}{T_{CMOS}} - \frac{1}{T_{dark}}\right)\right)}
\]

where $t_{exp}$ is the exposure time of the actual frame, $t_{expDark}$ the exposure time of the dark frame, $T_{CMOS}$ is the detector temperature of the actual frame, $T_{dark}$ the exposure temperature of the dark frame.

Images acquired in the thermal-vacuum chamber (TVAC) with the shortest exposure were examined for bad pixels. A median image was calculated from each temperature cubes.  No dead pixels were detected, but there are 7 "hot pixels", with DN values above the median (3 for AFC1, 4 for AFC2). These pixels are listed in Table \ref{tab:hot_px} and will be monitored during the mission. Pixels with 20 DNs larger than the frame average (approx. 17 DN) are considered suspicious.

\begin{table}[h]
\caption{Image coordinates (row, column) of hot pixels. (0,0) is the top left corner in the FITS files}\label{tab:hot_px}%
\begin{tabular}{l|l}
\toprule
AFC1        & AFC2 \\
row, column & row, column \\
\midrule
322, 490  & 923, 101 \\
645, 516  & 912, 152 \\
463, 775  & 828, 254 \\
          & 688, 866 \\
\botrule
\end{tabular}
\end{table}

\subsection{Flat Field tests}
The flat field tests utilised a 0.5 m diameter integrating sphere. The camera was mounted horizontally, with the optical axis facing the entrance aperture of the integrating sphere. The sphere was moved close to the camera baffle to avoid vignetting by the entrance aperture.
The sphere’s back surface was illuminated by halogen light sources, covering the full wavelength range of the AFC camera. The lamps’ current was gradually increased to 1.600 A. After 10 minutes, the lamps stabilised, and the flat images were acquired.
Three image cubes have been recorded with ten frames in each data file. The exposure times were set to 0.224 ms, 0.436 ms and 0.648 ms. This corresponds to approximately 25\%, 50\%, and 75\% maximum intensity in the central image regions.

The images show a uniformly illuminated field of view, with a slight (4-6\%) drop in intensity towards the edges.
The frames do not show signs of dust or other contamination of the detector or the optical components.
The two cameras have very similar flat images, and no sign of disturbing stray-light effects, vignetting, or internal reflections.

The normalised flat images are separated into a low-frequency component by applying a 200px Gaussian blur on it, and a
 high-frequency component by subtracting the low-frequency component. The normalised standard deviation of the high-frequency components is less than 0.005. The normalised standard deviation of the low-frequency components is less than 0.015.

The master flat frame for the calibration pipeline was created from the 0.436 ms exposure flat cubes, with the applied processing steps:
\begin{itemize}
\item The median image was calculated from the 10 cube frames.
\item The median was converted to a 64-bit PC double image.
\item The master bias image was subtracted from the median.
\item The image was normalised (scaled) to have the central 200x200 pixels average a value of 1.0.
\end{itemize}

\begin{figure}[h]
\centering
\includegraphics[width=\textwidth]{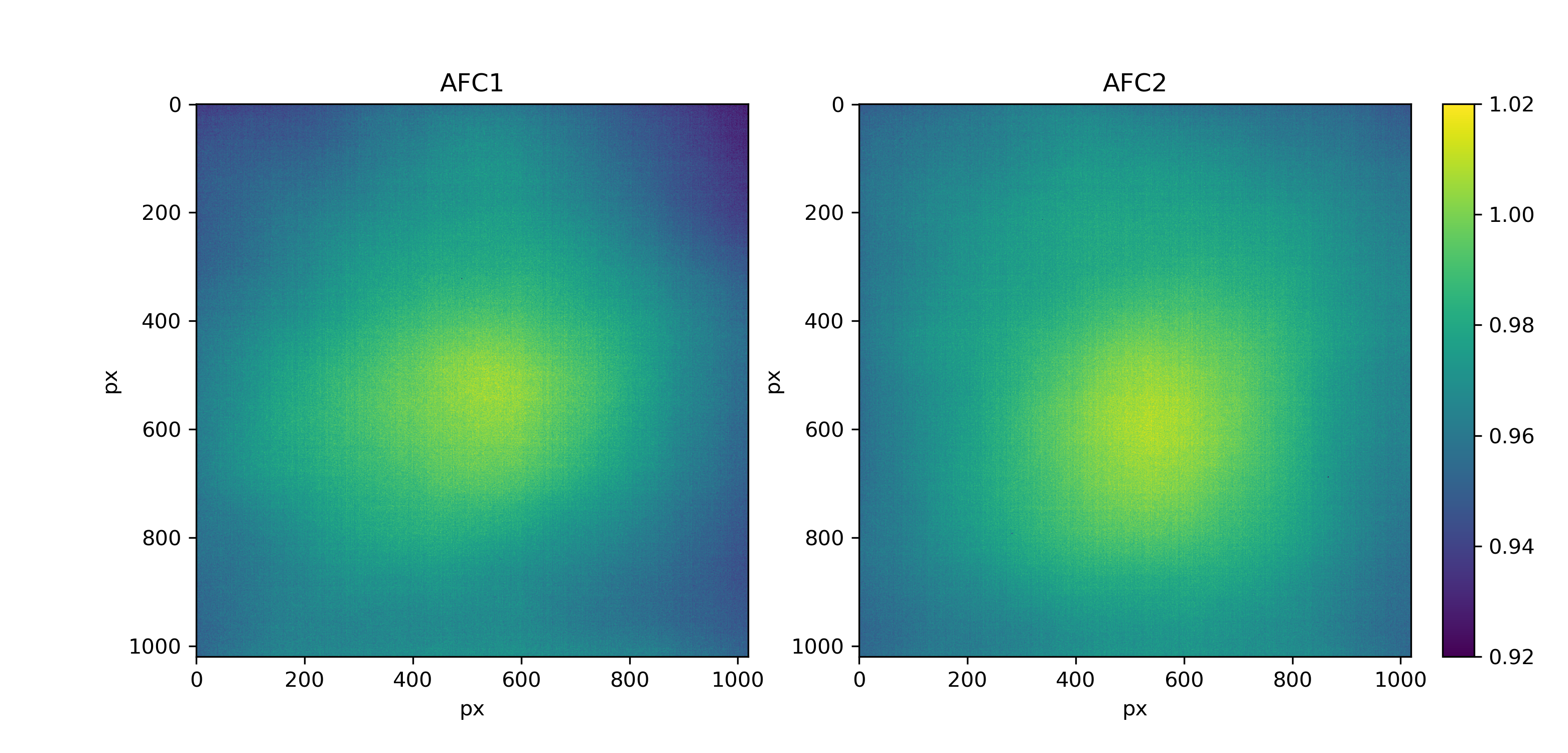}
\caption{Master Flat images for AFC1 (left) and AFC2 (right)}
\label{fig:flats}
\end{figure}

Figure \ref{fig:flats} shows the reference master flats for AFC1 and AFC2.

\subsection{Linearity}
The linearity test utilised the same optical setup as the flat field test. The camera was mounted horizontally, with the optical axis facing the entrance aperture of the 0.5m integrating sphere. The sphere was moved close to the camera baffle to avoid vignetting.
The sphere back surface was illuminated by halogen light sources, covering the full wavelength range of the AFC camera. The nominal sphere illumination was too high for the sensitivity of the camera and allowed only four exposure steps without saturation. For this reason, the sphere illumination was decreased by adjusting a lower lamp current than for the flat test.
After switching on, the current of the lamps was gradually increased to 1.150A. After 10 minutes, the lamps stabilised, and the linearity images were acquired. A total of 28 image cubes have been recorded, with three frames in each data file. The exposure time was gradually increased from 0.224ms to 5.951ms.

For each data cube (with the same exposure settings) we applied the following steps:
\begin{itemize}
\item The average image was calculated from the 3 cube frames.
\item The master bias image was subtracted from the median.
\item The image average pixel and standard deviation were calculated.
\item A linear fit was calculated from the averages and the exposure times and used as an estimate of the ideal pixel averages.
\item The linearity error was calculated as the difference between the actual and the ideal average.
\end{itemize}

This procedure checked the detector response linearity vs. exposure time, in the short exposure time regime. This exposure range will be used for most of the asteroid imaging (when the illuminated detector area is larger than a pixel size). The test exhibited relatively large errors for the very short exposures, and the linearity improved above the 1.0 ms exposure.
For long exposure imaging ($> 200 ms$), the specific characteristics of the APS detectors must be considered: in the case of several hundred milliseconds exposure, the full well capacity of the pixels is significantly reduced. For instance, 800ms exposure images would saturate around 3300 DNs, and longer exposure might further decrease the saturation level.
This needs special attention at a low light level long exposure commanding since the detector is no longer saturated by the ADC (at 4096 DNs), but by the detector full well, at a lower DN value. Essentially, at short exposure times, the ADC saturates before the detector reaches full-well. However, as the exposure time increases, the detector reaches full-well or saturates before the ADC saturates and the level at which the device saturates drops with increasing exposure time.

We note that this behaviour, although undesirable, has no consequence on the instrument fulfilment of the mission requirements. All asteroid images will be acquired at very short exposure (typically $< 10 ms$), an order of magnitude less than the exposure threshold at which the image saturation level starts to decrease.

\begin{figure}[h]
\centering
\includegraphics[width=\textwidth]{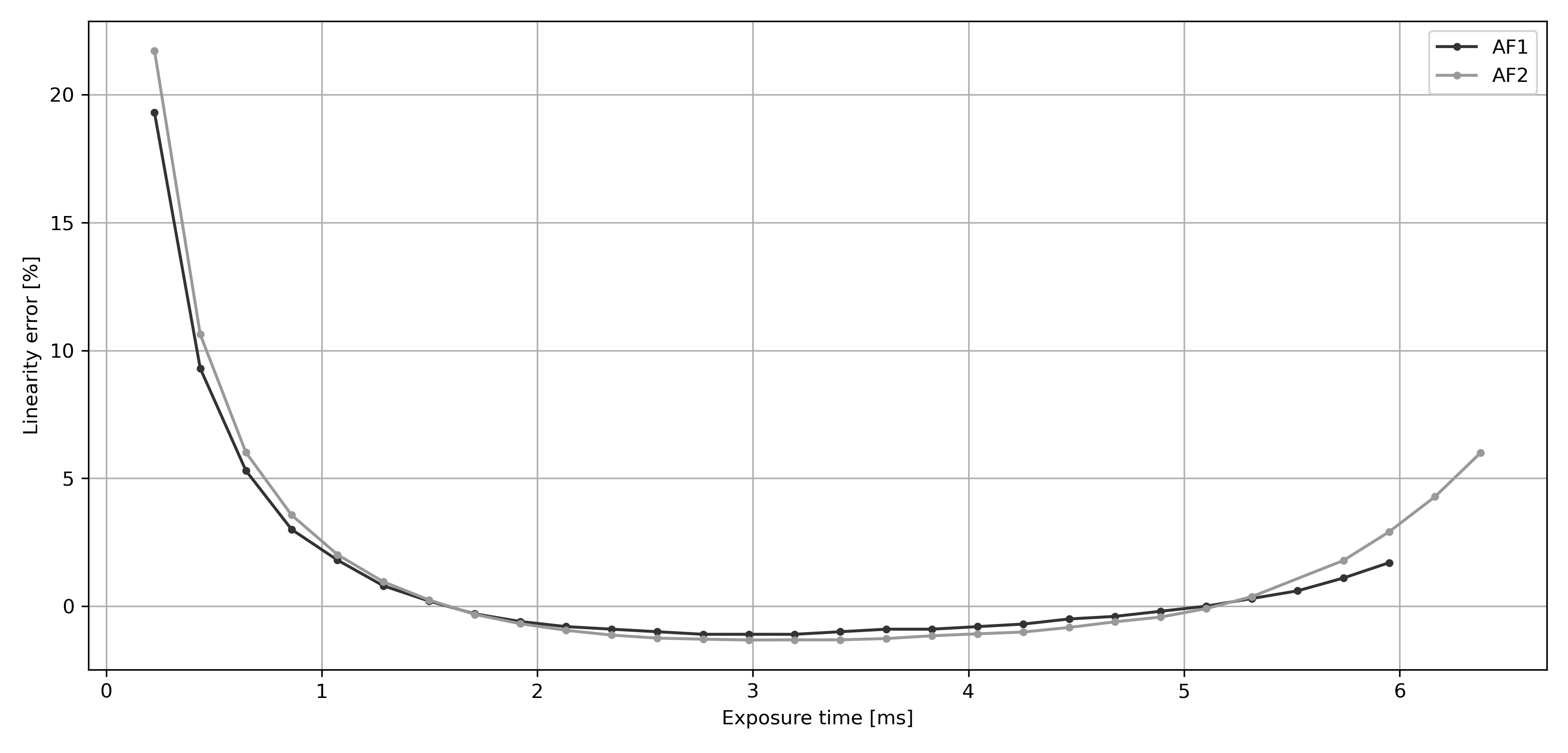}
\caption{Linearity error vs. exposure time at high intensities for both cameras .}
\label{fig:lintest}
\end{figure}

Figure \ref{fig:lintest} shows the results of the linearity measurements.

\subsection{Residual Image tests}
The goal of the residual image test is to confirm the effectiveness of the detector clear process, and to check for the existence of a residual image from the previous exposure. Since the detector does not provide a true 0.0ms exposure possibility (image readout without actual integration), the accurate determination of the residual image is difficult. The test utilised the same optical setup as the flat field test. The camera was mounted horizontally, with the optical axis facing the entrance aperture of the 0.5m integrating sphere. The sphere was moved close to the camera baffle to avoid vignetting.
The sphere’s back surface was illuminated by halogen light sources, covering the full wavelength range of the AFC.
The test commanded short exposure (0.224ms) and long exposure (5.525ms) image acquisition in a quick sequence, without wait time between them. The test was repeated four times.

After exposure, the bias removed images were visually examined for inconsistency and traces of residual DNs on the short exposure image. The ratio of the bias removed short and long exposure DN average was calculated.

The short exposure images did not show traces of residual image for any of the cameras, the average DNs correspond to the scaled long exposure image.

\subsection{Radiometric response}
The radiometric test measured the camera response for a variable, narrow-band (quasi-monochromatic) light source. The source comprised a strong halogen light source, a grating monochromator, and a small integrating sphere. The monochromator scanned through the visible and NIR spectrum from 400 to 1000nm, with 10 nm steps. The monochromator bandwidth was set to 10nm. The monochromator output was coupled into a small integrating sphere. The camera was imaging the uniform radiation of the integrating sphere output aperture (10mm diameter). Since the calibration target was not in focus, the image processing required integrating the flux in the ROI area. The actual output flux was monitored by a custom radiometer with calibrated Si diode.

To process the image cubes, the following steps were carried out:
\begin{itemize}
\item The median of each image cube was calculated.
\item The bias and dark were removed based on the master bias and the scaled master dark frame.
\item Pixels on the "bad pixel list" are discarded.
\item The camera response was calculated by aperture photometry. A circular aperture of 126 pixels in diameter was used to collect the monochromator flux, and a 50-pixel wide ring with a 0-pixel shift was used as the annulus.
\item The output of the monochromator was monitored by a silicon diode, and its photo-current was converted to radiance units.
\end{itemize}

\begin{figure}[h]
\centering
\includegraphics[width=\textwidth]{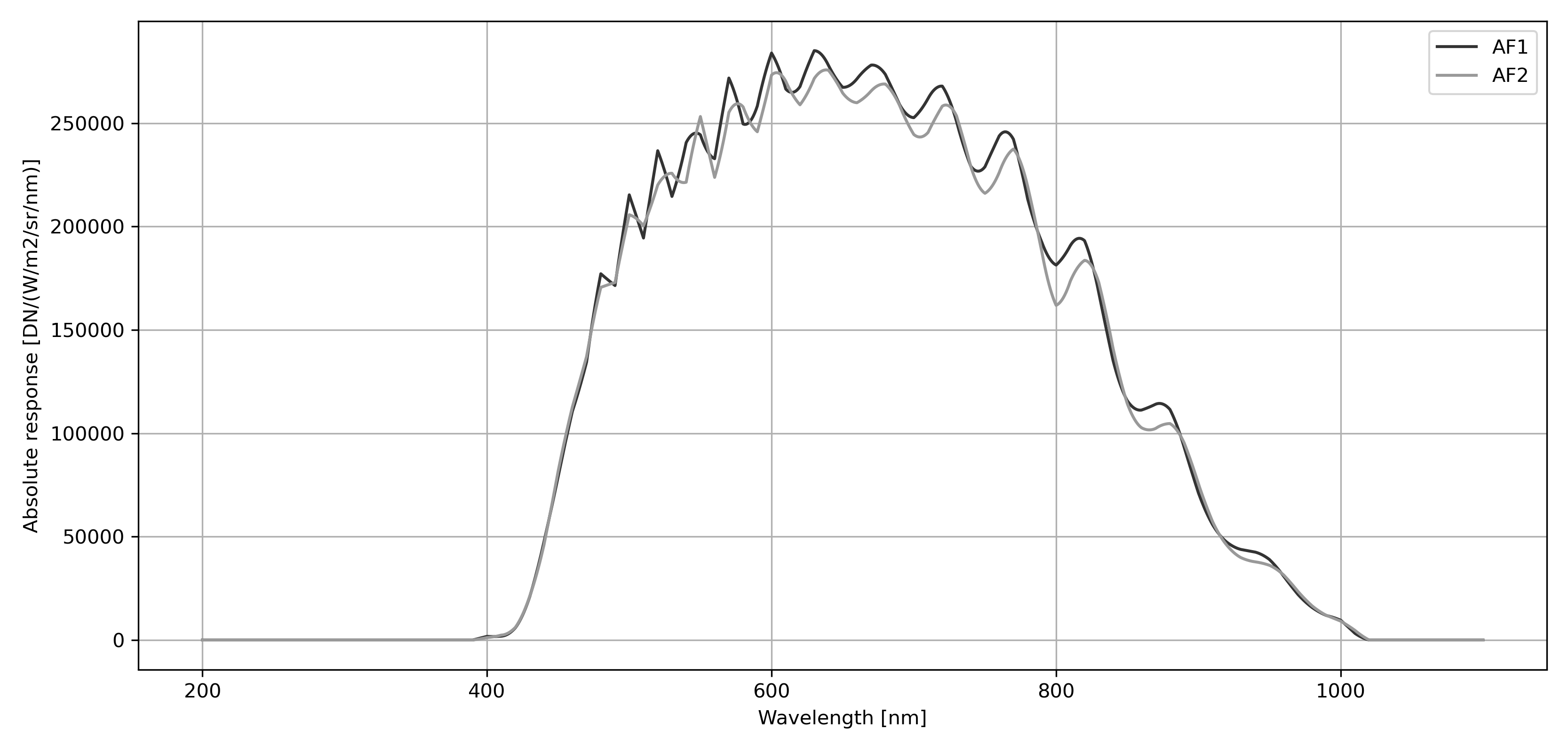}
\caption{Absolute response of the cameras as a function of wavelength.}
\label{fig:radtest}
\end{figure}

The two cameras provide a very similar radiometric response. The spectral characteristics match within 2\% (Fig. \ref{fig:radtest}).
The integral camera responses are higher than the theoretical calculations based on the detector and optical data. This might indicate a slightly higher electronic gain value than specified.
The spectral characteristics are also slightly different than expected from the detector specifications: Both cameras exhibit an oscillation of the response in the wavelength domain across the sensitive range. This might be the effect of some dichroic layers on the detector (or optics). Since the camera is acquiring only broadband monochromatic images, this does not create a significant problem; however, it decreases the photometric accuracy, especially for Solar illuminated targets.

In addition to the monochromatic tests, the camera radiometric response was also evaluated in the context of broad band imaging of an extended source, which is the actual mode relevant for all operations in the Didymos system. The test setup used a target lamp which had a circular aperture with Lambertian radiating surface. The radiance was calibrated in a certified calibration laboratory in Hungary.

The lamp was positioned at 4035 mm in front of the camera. The objects around the lamp and in the field of view of the camera were covered by black shading, to avoid stray light reflections. The lamp current was gradually increased to the nominal value. After the lamp stabilised, image cubes were acquired with correctly and overexposed settings. The overexposed images were intended to reveal possible stray light.

This setup is a good proxy for the asteroid approach phase when the target size is about 20\% of the full field.

After exposure:
\begin{itemize}
\item The median of the image cube was calculated.
\item The bias and dark were removed based on the master bias and the scaled master dark frame.
\item The mean DN value of the central 51 pixels diameter area was calculated.
\item The measured camera response was compared to the estimated values, based on the lamp radiance calibration.
\end{itemize}

We report that the expected radiance values match the lamp calibration measurements within 5\%.

\begin{figure}[h]
\centering
\includegraphics[width=0.45\textwidth]{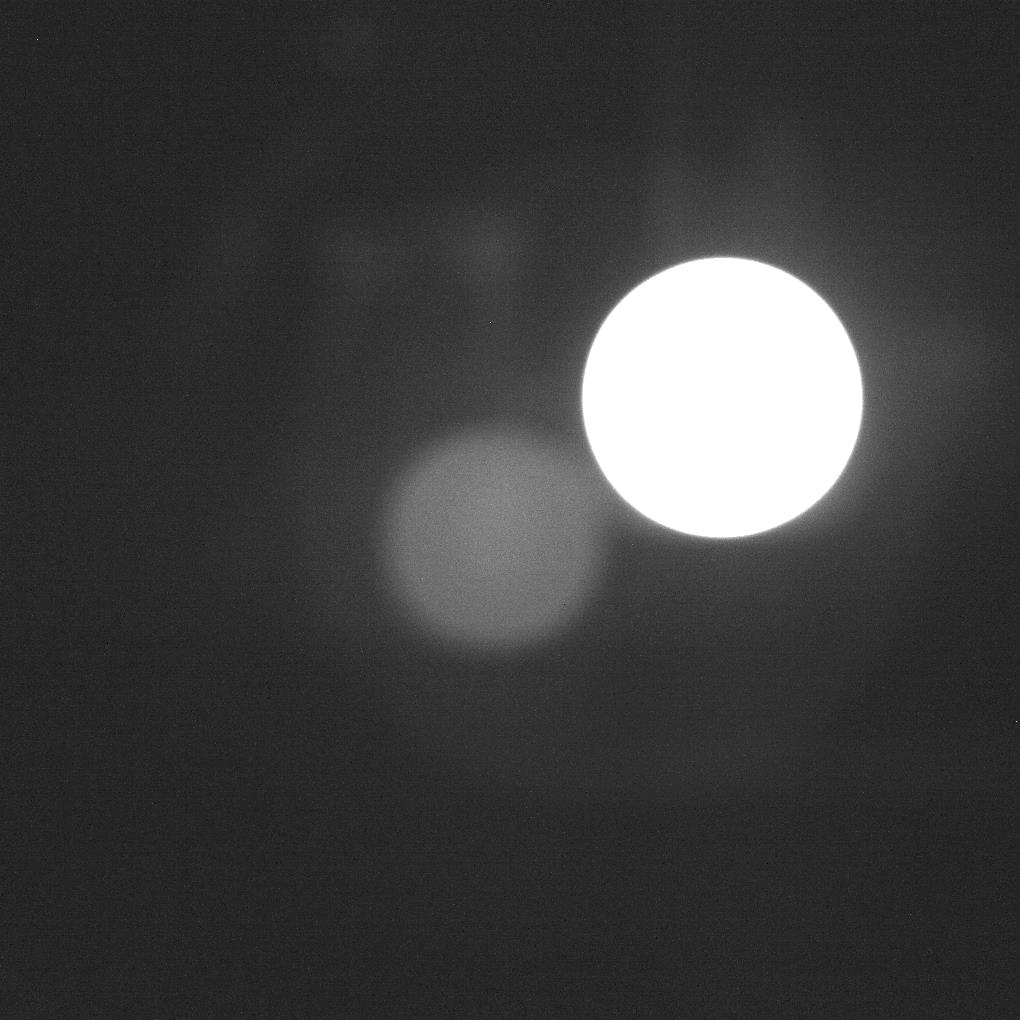}
\caption{Ghost pattern on the overexposed broad band image}
\label{fig:ghost}
\end{figure}

The broad band calibration lamp test also provided an opportunity to test for the presence of in-field stray light effects. When the lamp was not on the centre position, a slight ghost pattern was visible on all images in case of stretched intensity. The lamp position was also shifted by 150 mm, and overexposed images were also acquired. The intensity of the ghost pattern was very weak, less around 0.1\% of the lamp image pattern (Fig. \ref{fig:ghost}).

As a general recommendation, we calculated the camera reference wavelength from the camera DN response curves and the solar reference illumination, and measured the corresponding calibration factors. We recommend the following parameters:

\begin{table}[h]
\caption{Radiometric calibration factors}\label{tab:calfactor}%
\begin{tabular}{l|l|l}
\toprule
Parameter & Value & Unit\\
AFC1 effective wavelength & 654.7 & nm \\
AFC1 effective FWHM & 366.0 & nm \\
AFC2 effective wavelength & 655.6 & nm \\
AFC2 effective FWHM & 370.8 & nm \\
General reference wavelength & 650.0 & nm \\
Solar Irradiance (1 AU) & 1.5794 & $W.m^{-2}.nm^{-1}$ \\
Calib. Factor AFC1 & $9.5877\times10^5$ & $DN.m^2.nm.W^{-1}.s{-1}$ \\
Calib. Factor AFC2 & $9.3698\times10^5$ & $DN.m^2.nm.W^{-1}.s{-1}$ \\
\botrule
\end{tabular}
\caption{FWHM: Full Width, Half Maximum}
\end{table}

\subsection{Geometric Distortion}
\begin{figure}[h]
\centering
\includegraphics[width=\textwidth]{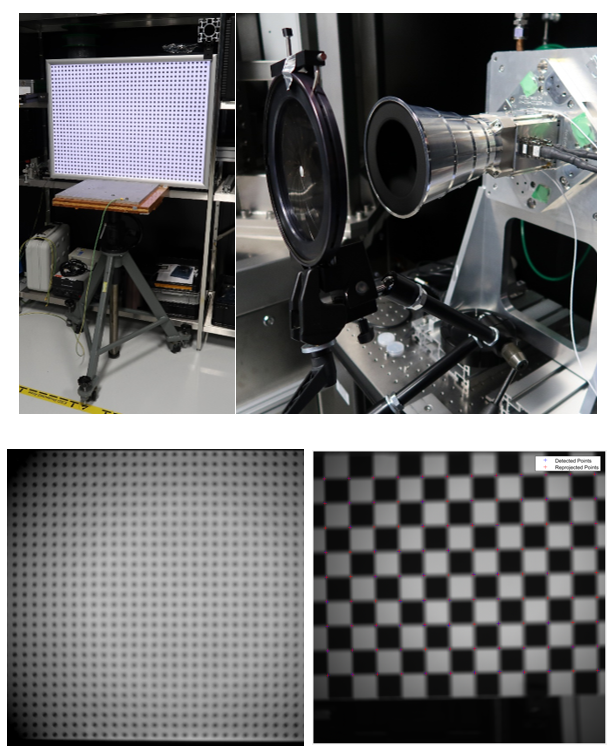}
\caption{Test setup for optical distortion measurements}
\label{fig:dist}
\end{figure}

The main objective of the test is to measure the camera geometric distortion, and to provide a correction function to resample the image with a constant magnification across the full field of view. The camera distortion tests were carried out in the JOP optical laboratory, at ambient conditions (P = 1013 hPa, T = 24°C). Since the image plane position is adjusted for vacuum, and the test target distance was limited by the laboratory space, the test images were not perfectly in focus. This fact has a limiting effect on the accuracy of the distortion maps. In order to improve the image distortion correction, star-field images will be acquired during the cruise from Earth to Didymos.
The ground tests were performed by three different targets:
\begin{itemize}
\item Black dots (10mm diameter) on 20mm pitch, self-illuminating cool white LED background
\item Black and white chequerboard, 10mm by 10mm squares, ambient illumination from the laboratory lights
\item Black and white chequerboard, 10mm by 50mm squares, ambient illumination from the laboratory lights
\end{itemize}
To improve the sharpness of the images, a circular aperture (20mm diameter) was placed in front of the camera (Fig. \ref{fig:dist}).

Both cameras produced very similar distortion images. No apparent assembly effects (component tilt/decenter) were found. The ambient air pressure and the finite object distance resulted a strongly defocused image. Also, the illumination variation across the targets caused intensity variation on the images. These effects were strongly influencing the accuracy of the control points. The positional accuracy of the control points was estimated to be 1.5 pixel (min 0.9 pixel, maximum 5 pixel).
 As the first approximation of the distortion correction only radial distortion components were considered. The two cameras were supposed to behave similarly, so the radial distortion fit was calculated for a unified shift dataset.

The radial distortion is provided as a degree-3 polynomial function:
$$
D_{rad} = K_0 + K_1 r + K_2 r^2 + K_3 r^3
$$
where $D_{rad}$ is the radial distortion in mm, $K_0...K_3$ are the polynomial coefficients (Table ), and $r$ the radial distance of the image point, measured from the geometrical centre of the image: [509.5, 509.5] pixels

\begin{table}[h]
\caption{Polynomial coefficients for the radial distortion}\label{tab:hot_px}%
\begin{tabular}{l|l|l|l}
\toprule
$K_0$ & $K_1$ & $K_2$ & $K_3$ \\
\midrule
$0$ & $-1.1398\times 10^{-4}$ & $9.0621\times 10^{-5}$ & $-1.3891\times 10^{-4}$ \\
\botrule
\end{tabular}
\end{table}

\textit{Note that the distortion will be better evaluated in flight through imaging of reference star fields. Preliminary analysis suggests that the distortion is minimal: maximum error = 0.81 pixel, average error = 0.33 pixel.}

\subsection{Point Spread Function}
The point spread function (PSF) was not measured on the ground, but instead calculated from many images of star fields acquired in the first 6 months of cruise.
On star field images the average PSF size is less than 2 pixels including the pointing instability. The Gauss fit S parameter is about 0.8 pixel (1.6 pixel diameter) which encircles 65\% of the star flux.

\subsection{Recommended minimal procedure for image calibration}
The suggested minimal AFC flight calibration pipe should implement the following steps:
\begin{enumerate}
\item Open image file(s): check data integrity, extract header metadata, extract binary content.
\item Convert image pixel 16-bit DNs to double precision numbers.
\item Remove bias based on the master bias frame.
\item Remove bad pixels.
\item Correct for dark current using the master dark image and the dark scaling equation with the header information (exposure time, detector temperature). Master dark images shall be updated during the mission.
\item Apply the flat field correction.
\item Correct linearity errors.
\item Scale the image to radiance or I/f (reflectance) units.
\item Apply distortion correction (if needed for accurate morphology or navigation)
\end{enumerate}

\subsection{In-flight Calibration}
To monitor the state of the instruments, we designed a set of calibration sequences to be acquired at regular intervals in flight. The sequences include monitoring of the dark and bias levels, observations of two reference star fields for measuring geometric distortions, and two standard stars for radiometric calibration. All targets are listed in Table \ref{tab:calib_targets}. All observations will be acquired with multiple exposure times. These calibration activities will take place every 6 months during the cruise. After arrival at Didymos, we will monitor the dark and bias values on a weekly basis, and point to stars and star fields for radio/geo-metric calibration once per mission phase (see Section \ref{sec:ast}).

\begin{table}[h]
\caption{Calibration targets}\label{tab:calib_targets}%
\begin{tabular}{l|l|l|l}
\toprule
Target        & RA      & DEC      \\
\midrule
Starfield & 07h 21m & -25d 34m \\
Starfield & 04h 25m & +15d 10m \\
Vega      & 18h 37m & +38d 47m \\
16 Cygni  & 19h 42m & +50d 31m \\
\botrule
\end{tabular}
\end{table}

During cruise, Hera will have two unique opportunities to observe extended objects: the Earth/Moon system shortly after launch, and Mars/Deimos in March 2025.

The Earth/Moon images will be acquired at high phase angle ($>75\degree$) a few days after launch. We expect the Earth to occupy at least 50 px, and the Moon at least 15 px in the AFC images, but the exact size depends on the spacecraft status after launch and how early the observations can be scheduled.

\begin{figure}[h]
\centering
\includegraphics[width=\textwidth]{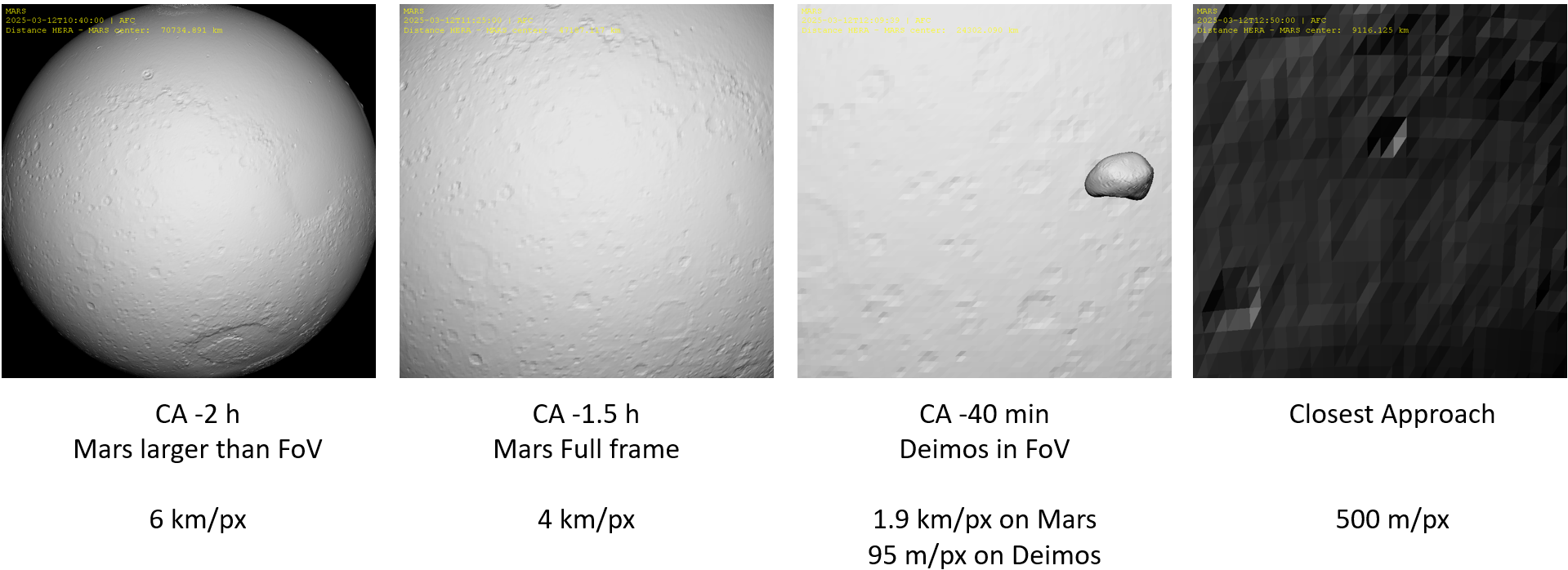}
\caption{Simulated AFC observations showing the expected views and pixel scale during the Mars swing-by. Exact time and distance of closest approach are preliminary, pending results of trajectory correction manoeuvres on the way to Mars. The simulation only shows the geometry of shape models, with a simple Lambert shading. This is not intended to be radiometrically correct.}
\label{fig:mars}
\end{figure}

The Mars swing-by will allow for observations of a full rotation of Mars at varying scale, including
many frame filling images. During the swing-by, Hera will cover a wide range of phase angles ($0\degree - 180\degree$), providing a excellent data set for radiometric calibration of the full detector. High-resolution images of well-known terrains will be used to refine our measurements of the geometric distortion.
In addition, Hera's swing-by is timed so that Deimos will be in the field of view of the AFC, in between Hera and Mars, from a distance of 1000 km. This will allow to resolve a spatial scale of 95 m/px on the far side of Deimos (1.9 km/px on Mars in the same image). Figure \ref{fig:mars} shows simulated views of the swing-by.

Although this article focuses on the AFC, it should be noted that we strive to acquire the best data for cross-calibration between all remote-sensing instruments on Hera (AFC, Hyperscout, TIRI). As much as possible, we align our calibration observations to target the same objects at the same time (stars, star fields, planets).

\section{Asteroid Phase}\label{sec:ast}
The cruise phase activities are limited to commissioning and calibration activities, with the exception of the "farewell to Earth" images, and Deimos far side observations.

As we approach the asteroids, the AFC will perform a reconnaissance of the system with deep exposures to look for any eventual debris that could be hazardous to the mission. We will also acquire light curves to refine our understanding of the shape and spin properties of both asteroids.

The core AFC activities will take place at the Didymos system from December 2026 and last for at least 6 months. The mission is divided in 5 major phases which will bring Hera closer and closer to the asteroids, achieving full mapping of the two bodies at increasingly better resolution.
\begin{table}[h]
\caption{Mission phases}\label{tab:mission_phases}%
\begin{tabular}{l|l|l}
\toprule
Phase                                & Duration (weeks) & Distance (km) \\
\midrule
ECP: Early Characterization Phase    & 6 & 30 - 20 \\
PDP: Payload Deployment Phase        & 2 & 30 - 20 \\
DCP: Detailed Characterization Phase & 4 & 20 - 10 \\
COP: Close Operations Phase          & 6 & 20 - 5  \\
COP: EXperimental Phase              & 4 & 20 - 1  \\
\botrule
\end{tabular}
\end{table}

\begin{figure}[h]
\centering
\includegraphics[width=\textwidth]{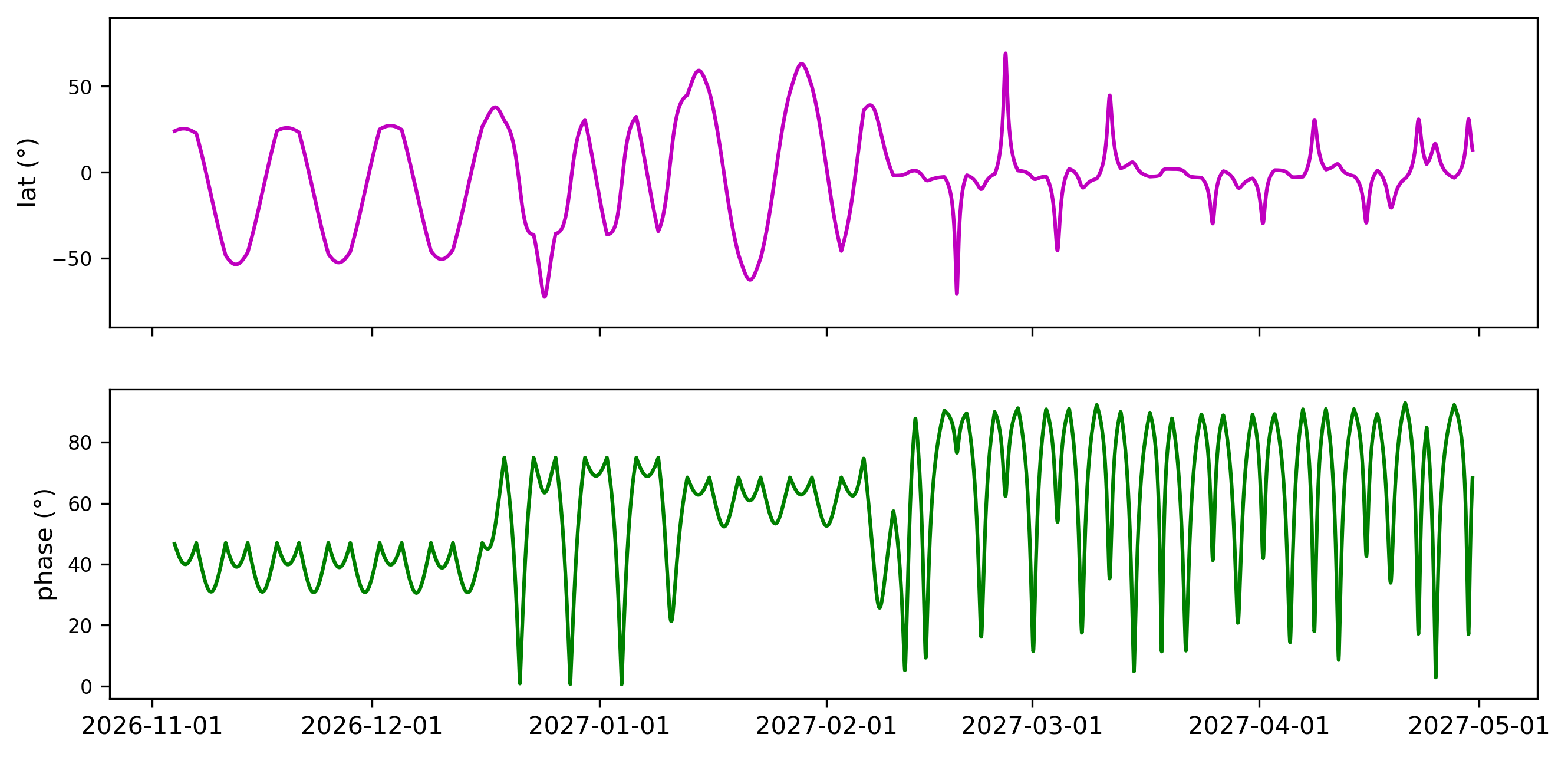}
\caption{Sub-spacecraft latitude and phase angle on Didymos/Dimorphos in the first four phases of the mission after arrival at Didymos. All dates are preliminary and may change during cruise.}
\label{fig:angles}
\end{figure}

At the time of writing this article, the detailed trajectory of Hera in the Didymos system has only been fully evaluated until the end of DCP. We provide in Figures \ref{fig:dist} and \ref{fig:angles} the planned distances, resolution, latitude and phase angle coverage in the first three phases in the mission. These shows that Hera will observe the full surface of both asteroids, except perhaps polar areas that may be permanently shadowed. The phase angle will be mostly between $40\degree$ and $60\degree$, which provide ideal viewing conditions for mapping and shape reconstruction. In DCP, the trajectory provides 4 opportunities to observe the surface at zero phase from a distance of 10 km (i.e. 1 m/px resolution), which is intended to provide albedo measurements of both asteroids.

\begin{figure}[h]
\centering
\includegraphics[width=\textwidth]{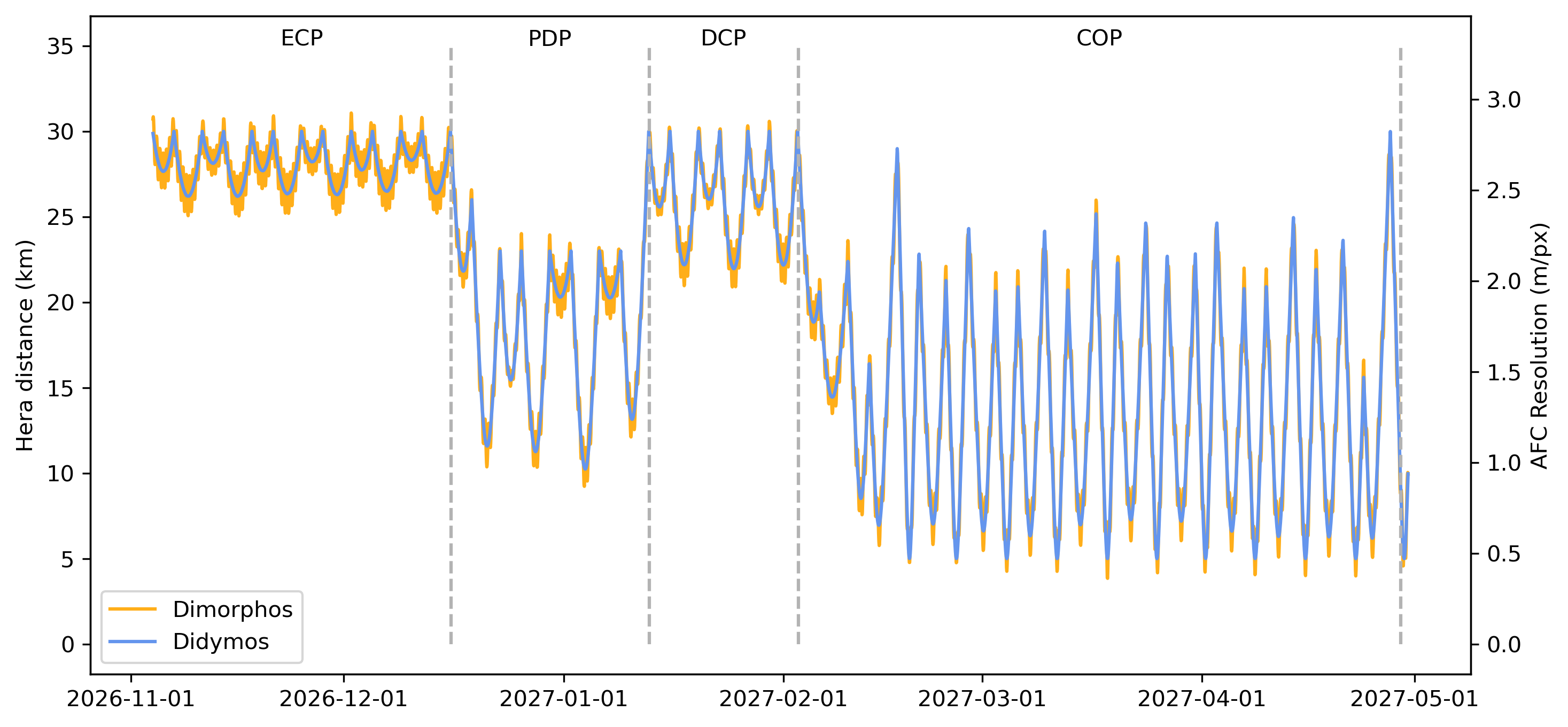}
\caption{Planned trajectory for the first four phases of the mission after arrival at Didymos, ECP: Early Characterization Phase, PDP: Payload Deployment Phase, DCP: Detailed Characterization Phase, COP: Close Operations Phase. All dates are preliminary and may change during cruise.}
\label{fig:ops}
\end{figure}

During the mission, the AFC mostly operates in a mapping mode. We want to ensure consistent data set with similar observations that repeat day after day in order to maximise the chance to detect changes, whether as surface transformation or dynamical evolution of the system. To achieve this, our baseline is to acquire one image every 10 min while in the Didymos system, with a duty cycle of 15 hours per day (+ 9 hours of downlink/manoeuvres). This ensures that Dimorphos will be observed for more than a full orbit every day (current orbital period ~11.5 hours, \cite{thomas2023}). The mapping cadence may be reduced to 20 min at the end of the mission, when data volume might be more constrained.

This mapping data will be used for shape determination, orbital state, dynamics, morphological mapping, identification of DART crater and other features of interest, selection of fly-bys targets (in preparation of COP), change detection.
Once per week, we plan some deep exposures and fast cadence imaging to detect eventual dust emission or track potential debris in the system. These observations may be more frequent if needed.

Finally, we will make use of the synergy with other instruments to increase the science return of the mission. For instance:
\begin{itemize}
    \item the AFC will track Dimorphos as it enters the shadow of Didymos to provide high-resolution context to the thermal imaging instrument (TIRI, \cite{deleon2024}). Similar joint observations will be acquired when tracking the shadow of Dimorphos cast onto Didymos.
    \item Global mappings of the asteroids with the Hyperscout instrument \citep{okada2024}, at low phase angle, will allow to cross-calibrate our albedo measurements.
    \item At altitudes below 17 km, we will operate the AFC together with the laser altimeter PALT, in order to combine the ranging and imaging data to build better shape models of the asteroids.
    \item We will characterise the landing sites of Juventas and eventually Milani at high resolution, before and after landing, to better understand the response of the surface to the landing event.
    \item If the trajectory allows it, we plan to track the probes as they are released. Observing their landings as they occur might be challenging but possible if Hera is $< 1 km$ from the cubesats.
\end{itemize}

Overall, the data to be acquired by the AFC amount to about 20 000 images, typically 90 exposures per day (135 MByte) in the asteroid phase.

\section{Scientific activities}\label{sec:sci_activities}

ESA's Hera mission will bring the key measurements necessary to close the AIDA experiment: what is the actual momentum transferred by the impact (mass), what is the final state of Dimorphos? (crater, reshaped), what are the asteroids internal properties (rubble pile, monolith).

It is also a mission of many "firsts" for asteroid science, as detailed in \cite{michel2022, michel2024}:
\begin{itemize}
\item First binary asteroid orbited by a space mission,
\item Smallest asteroid ever visited by a space mission,
\item First internal structure probing of a full asteroid,
\item First asteroid visited near the disruption spin barrier.
\end{itemize}

These questions cannot be answered without the contributions from an instrument like the AFC, as imaging data is a necessary requisite for most of the studies carried out by the investigation team.

We have presented the technical specifications of the camera, and the calibration results, which gives confidence that the instrument is perfectly suited for its tasks. In parallel, the team has been developing many tools to analyse the data: from shape reconstruction pipelines \citep{preusker2017}, to automated image processing for boulder statistics \citep{pajola2024} and morphology \citep{robin2024}, roughness analysis \citep{vincent2023}, as well as 3d visualisation software (\cite{vincent2018, caballo2024}).

~\\
A key component of the science to be enabled by the AFC is the production of a shape model for each asteroid. This data product will be the reference for all studies that need to know the volume of the asteroids (e.g. to derive density and porosity) and for morphological studies.

While images will be continuously acquired during the mission, and will be used for shape reconstruction, the first two phases of the mission (ECP: Early Characterization Phase, and DCP: Detailed Characterization Phase) are particularly well suited for stereo-photogrammetric analysis. In both phases we currently aim to acquire images with a 10 minutes cadence.
To assess the expected quality of the shape reconstruction, we performed a study of the viewing conditions. Indeed, the stereo-photogrammetric analyses require a favourable image- and illumination geometry, because such geometry affects the quantity and quality of matched image points and resulting reconstructed surface points. Furthermore, the stereo constellation (stereo angle and number of stereo partners) influences the reconstruction result, where only a constellation with at least three stereo images results to a stable reconstruction of a surface point. Due to the rotational properties of both bodies the observation conditions vary during these observation phases, affecting the quality of stereo conditions accordingly. Table \ref{tab:SPG} defines two different parameter sets of stereo requirements which are intended to enable complete stereo coverage in gradations (Table \ref{tab:SPG}).

\begin{figure}[h]
\centering
\includegraphics[width=\textwidth]{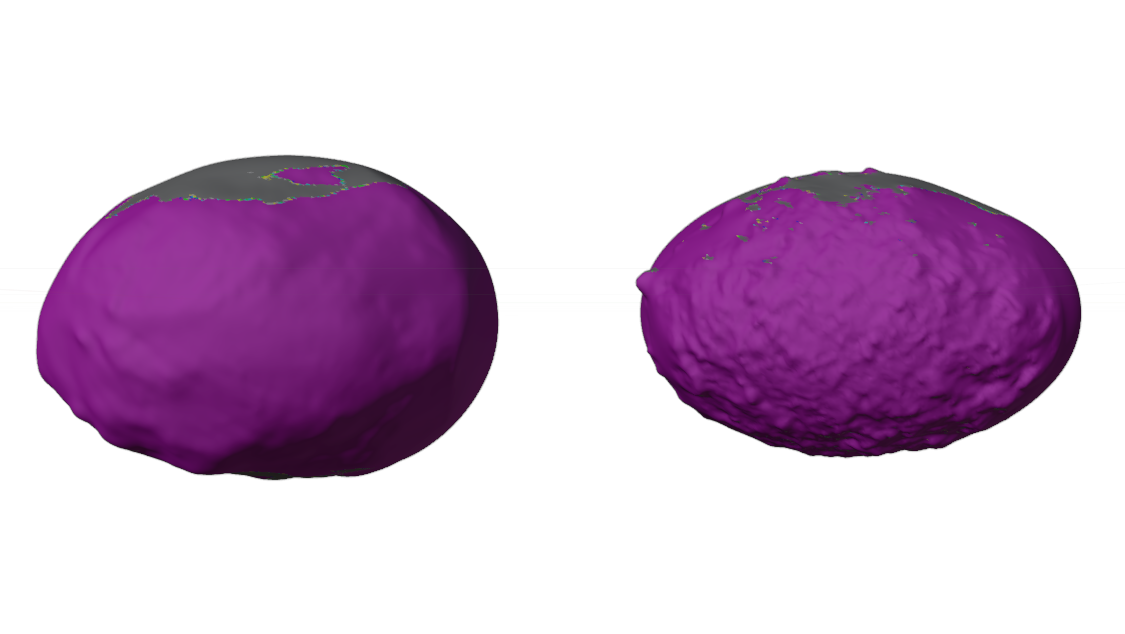}
\caption{Using our requirements, stereo coverage maps were calculated for Didymos (left) and Dimorphos (right) for the respective observation phases, whereby the number of images of a stereo constellation is marked in different colors. Here, we show the reuslts for the optimal case in ECP. Green (three views), blue (four views), and purple colors (five and more views) indicate sufficient stereo conditions, whereas insufficient stereo conditions are marked with yellow and grey colors. Shape models oriented with North up, asteroids not to scale.}
\label{fig:SPG_coverage}
\end{figure}

Using the respective stereo requirements, stereo coverage maps were calculated for Didymos and Dimorphos for the respective observation phases (Fig. \ref{fig:SPG_coverage}). Our simulations of the observing conditions show that the trajectory and attitude planned for ECP and DCP enable an excellent stereo coverage of all regions, on both asteroids, except for a few limited areas that may be permanently shadowed (Table \ref{tab:SPG2}). Although we recommend an imaging cadence of 10 min, a more relaxed cadence of 20 min, for 15 hours/day (accounting for downlink windows) appears sufficient to meet our goals.

\begin{table}[h]
\caption{Requirements for stereo-photogrammetry}\label{tab:SPG}%
\begin{tabular}{l|l|l}
\toprule
Parameter               & Optimal	  & Minimal     \\
\midrule
Illumination variation	& 0°–10°      &	0°–10°      \\
Stereo angle	        & 10°–70°	  & 15°–70°     \\
Incidence angle	        & 5°–65°      & 5°–85°      \\
Emission angle	        & 0°–65°	  & 0°–65°      \\
Sun phase angle	        & 10°–170°	  & 10°–170°    \\
\botrule
\end{tabular}
\end{table}

\begin{table}[h]
\caption{Summary of SPG feasability study. Surface area for which at least 5 stereo views meeting the "Minimal" or "Optimal" requirements will be obtained.}\label{tab:SPG2}%
\begin{tabular}{l|l|l}
\toprule
Mission Phase & Didymos coverage & Dimorphos coverage     \\
\midrule
ECP	          & 97.1\% (minimal) & 98.0\% (minimal)      \\
  	         & 81.3\% (optimal) & 80.3\% (optimal)      \\
\midrule
DCP	          & 98.6\% (minimal) & 93.5\% (minimal)      \\
  	         & 84.6\% (optimal) & 78.0\% (optimal)      \\
\botrule
\end{tabular}
\end{table}

~\\
All the operations planning and analysis effort is coordinated by the instrument team, and the Hera Data Analysis Working Group \citep{michel2022}, which ensure that the AFC data and derived high-level products will be available to the whole team in a timely manner.

\section{Conclusion}\label{sec:conclusion}
\begin{figure}[h]
\centering
\includegraphics[width=\textwidth]{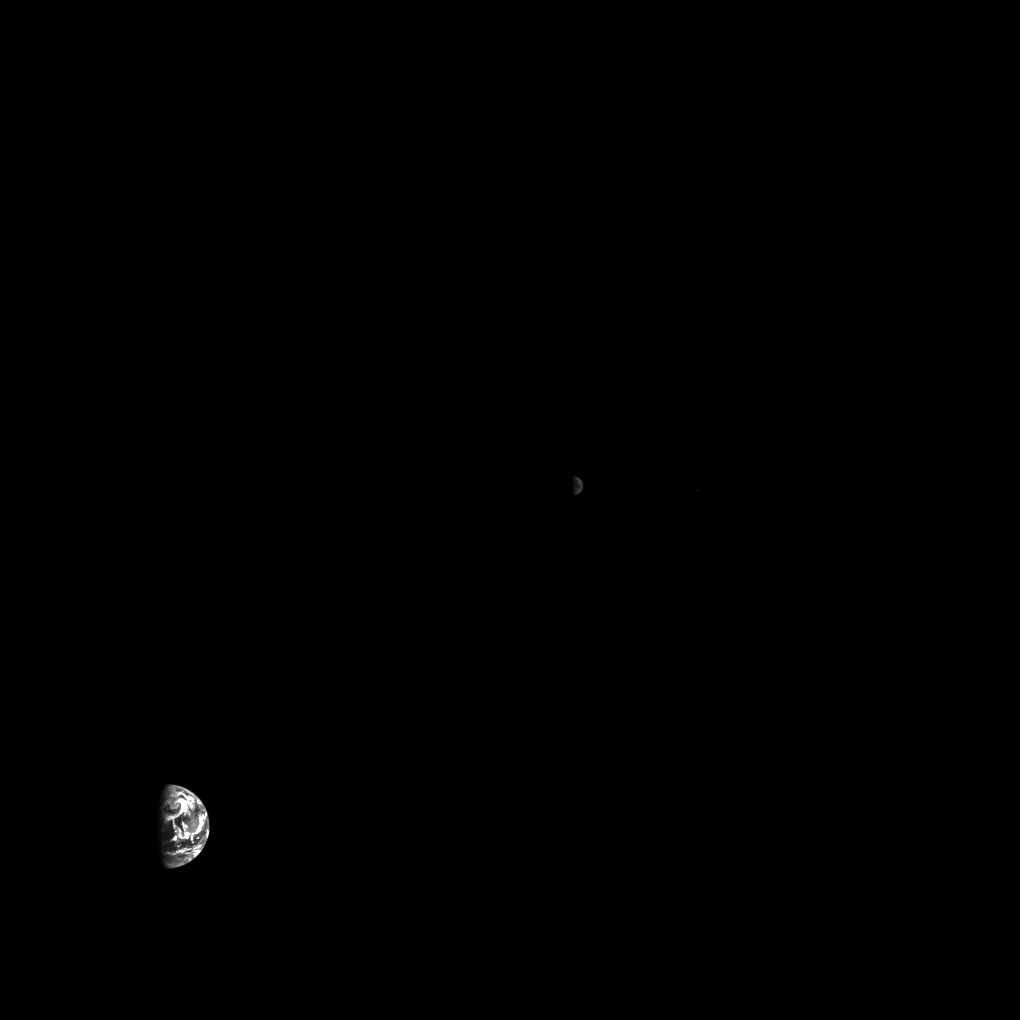}
\caption{First image acquired in space by AFC1 on 2024-10-11T21:25:06 UTC, showing the Earth (bottom left) and the Moon (centre) from a distance of approximately 1.6 million km. Earth is oriented with north pointing upwards, with the Pacific Ocean illuminated by the Sun.}
\label{fig:AFC_first}
\end{figure}

At the time of writing this paper, Hera has successfully launched and is on its way to Didymos. Instruments are being commissioned, and the AFCs have been switched on. Images of the Earth and the Moon have been acquired for calibration purpose (Fig. \ref{fig:AFC_first}), and show excellent behaviour from both cameras. A detailed calibration campaign will take place over the first 6 months of the mission, including the Mars swing-by. Results from this in-flight calibration will be presented in a future publication.

Beyond Planetary Defence, Hera provides a unique opportunity to observe in-situ a numerous but poorly known class of asteroids (binaries represent 15\% of Near-Earth objects). The AFC data will be instrumental in assessing the dynamical state of the system and predict its long term evolution. As for surface properties, the DART mission has provided a first view of the system pre-impact \citep{barnouin2024} which covers, however, only a fraction of the surface of both asteroids (at ~3.5m/px for Didymos, and 35cm/px for the global view of Dimorphos). While extremely useful, this data do not give us the complete picture, and we are missing global statistics on key morphological features such as craters and boulders, necessary to fully assess the age and evolution state of the surface. Global mapping with the AFC at multiple resolutions, will allow us to refine these statistics and build a complete picture, post-DART.

Indeed, it is a certainty that Dimorphos has changed because of the impact, and the AFC data will identify whether DART left a crater, or reshaped the body \citep{raducan2024}. And if the latter, to what extent ? nevertheless Didymos too may display significant changes as the result of ejected boulders re-impacting its surface. With a rotation period of only 2.25 h, the asteroid is very close to the theoretical disruption limit of a cohesion-less rubble pile \citep{agrusa2024}. As evidenced by several studies \citep{barnouin2024, bigot2024, vincent2024}, the equatorial region of Didymos is smoother, perhaps less cohesive, than other areas. It is likely that ejecta from Dimorphos may have fallen back on Didymos, and triggered landslides or mobilization of the surface. Such changes will be identifiable in AFC images, and the continuous mapping provided by the instrument will allow us to monitor the surface for any subsequent activity once Hera reaches Didymos.

The Hera mission will complete the AIDA experiment. By building a comprehensive characterisation of a Near-Earth Asteroids, Hera will give us the knowledge to defend against a potentially hazardous object, when the time comes. The Asteroid Framing Cameras, a core payload of the mission, have been thoroughly tested and calibrated before launch and are ready to support the mission in achieving its goals.

\section{Declarations}
\subsection{Funding}
J.-B. V., P. M., G. K., B. V. N., and F. P. acknowledge support from ESA through contract 4000144275/24/NL/GLC

M.P. acknowledges support from the Italian Space Agency (ASI) through ASI-INAF Agreement No. 2022-8-HH.0.

\subsection{Conflict of interest statement}
Not applicable.

\subsection{Acknowledgements}
The AFCs have been calibrated at the optical laboratory of Jena-Optronik GmbH (JOP). During the calibration tests the calibration process have been supervised by Hannah Goldberg (ESA) and Axel Kwiatkowski (JOP).

The calibration procedures were designed by Gábor Kovács and Balazs Vince Nagy (BME MOGI). The instruments have been supervised and operated by JOP engineers Axel Kwiatkowski and Karla Rossmann.

The calibration setups and the devices have been operated by MOGI.

\bibliography{2024_SSR_AFC_biblio}

\end{document}